# IPPOG – Bridging the gap between science education at school and modern scientific research

Barbora Bruant Gulejova (on behalf of the IPPOG Collaboration)

# Abstract

The International Particle Physics Outreach Group (IPPOG) has been making concerted and systematic efforts to present and popularise particle physics across all audiences and age groups since 1997. Today the scientific community has in IPPOG a strategic pillar in fostering long-term, sustainable support for fundamental research around the world. One of the main tools IPPOG has been offering to the scientific community, teachers and educators for almost 10 years is the Resource Database (RDB), an online platform containing a collection of high quality engaging education and outreach materials in particle physics and related sciences.

# 1. Introduction

Main challenges the scientific community faces currently are the threat to the financial support of large experimental endeavours and falling interest of young people to engage in studies of STEM (science, technology, engineering and mathematics), especially physics.

In general physics is less popular than other natural sciences, suffering a stigma of being a science that is very mathematical, abstract, and complicated. It is perceived as disconnected from real world. Most of the students who claim to love science do not mean physics, not realizing how fun physics can be.

It is important to break these stereotypes based on the misperception of physics in society. Physics is a pillar of all natural sciences. Physics is about understanding the basic laws of nature, the world we live in. It explains how the world around us and within us functions. This fundamental definition of physics does not often come across to students from their physics school curriculum.

Solving physics problems satisfies curiosity, develops analytical thinking and promotes an inquiring mindset. Many problems require creativity and out-of-the-box thinking to be mastered and this leads to a set of skills useful in all aspects of life. Moreover, people with a background in physics and related STEM subjects are highly sought after in many professions, including business and industry. These skills provide a solid basis for a successful career.

Today science, technology and innovation are among the most powerful forces driving social change and development. Natural sciences are also at the heart of producing new solutions to many of the challenges posed, for example, by the increasing risk of climate change, poor water management, or misguided applications of new technology. Therefore, physics and related sciences play an important role also in achieving the Sustainable Development Goals of UN Agenda 2030 [1]. A new generation of STEM specialists must prepare innovative solutions for tomorrow, one in which both genders have an equally important role to play. Moreover, future particle physics / high-energy physics projects will require a long-term, worldwide commitment of significant monetary resources and human expertise.

Today in Europe less than 20% of the young people choose STEM studies, while the number of STEM jobs grows three times faster than any other job. If we do not change something, there will be 7 million new European jobs in STEM by 2025 and not enough people to fill them [2].

A nation-wide German survey in 2000 [3] has revealed that almost 70% of the students drop physics at the earliest possible opportunity, which in some parts of the country already happens at an age of 15 to 16 years. Why is that? Let's start our analysis by stating that the only modern physics topic students are confronted with by that age is the physics of nuclear reactors.

## 1.1. Rare exposure of society to modern physics

We live in a modern world surrounded by technologies resulting from the greatest scientific discoveries of the past decades that are underpinned by modern physics of 20th and 21st centuries. The cell phones and computers upon which we rely were developed thanks to a fundamental understanding of the Standard Model of Particle Physics. Despite this, the physics taught in most of the high school classrooms is stuck in the past. "I always found it dry and removed from real world, but I learned that physics can be extremely fascinating and relevant, and I now realized that I actually enjoy physics, just not what is taught in our school curriculum.", claims a 16-years old participant of the programme 'Creating Ambassadors for Science in Society' [4] organized under the umbrella of the International Particle Physics Outreach Group (IPPOG) [5].

The main reference for physics teachers around the world today, the Halliday and Resnick collection known as Fundamentals of Physics [6], dates from the late 1950s [7]. We teach knowledge from centuries ago, and yet the questions, which inspire great fascination in youngsters (and even all of us!) are explained by contemporary physics, in particular particle physics, astrophysics, cosmology and quantum physics.

Perception of physics by students today is very different from the one of Marie Curie: "I am among those who think that science has a great beauty. A scientist in his laboratory is not only a technician: he is also a child placed before natural phenomena which impress him like a fairy tale."

## 1.2. Why understanding of particle physics by a broad population matters

Every child is a natural scientist, very curious with desire to learn. Kids often ask fundamental questions, such as "what are we made from?", "what is the universe made of?", "what is its origin?", "how has it come to be what it is today?", "what will be its end?", "why is the sun shining?"... Such questions have made human beings think about how the universe works and impel us to try to unveil its origins and evolution, and by investigating them science has advanced.

All around us and everywhere in the universe, particles bind, separate, are transformed or disintegrate. This is why the moon revolves around the Earth, the stars shine, magnets attract or repel each other, objects fall etc. Particles and their interactions are at the origin of our existence. Particle physics brings insights into the tiniest dimensions of life. It enables us to acquire deeper knowledge about the structure of matter, and hence, of the world around us. Particle physicists ask the biggest scientific questions imaginable and then devise ways to answer those questions.

The Standard Model of particle physics [8] is the current best theory for the fundamental structure of matter. It describes the results of essentially all terrestrial experiments in the physics of the microworld to a very high level of precision. Even though it is not yet complete and in quest for a more complete theory physicists work on understanding neutrinos, dark matter, antimatter and more, there is no justification to cloak particle physics in aura of mystery and to explain it only to a handful of privileged students.

Almost 50 years after the Standard Model was developed, in typical curricula worldwide physics lessons often take the form of a history class with live demos and occasional hands-on experiments. This gives an impression that physics' greatest discoveries were in the past. Most of the students end up finishing high school believing that there are only 3 elementary particles (electrons, protons, neutrons) and only knowing 2 types of forces (gravitational and electromagnetic) all as classically described. They never learn that the electron has five other sibling particles - electron neutrino, muon, muon neutrino, tau, and tau neutrino - which are collectively called leptons. They never learn that everything in the universe, from the cereal they eat for breakfast to the galaxies above, is fundamentally composed of just two quarks and one lepton. They will never learn that everything has the same origin and we are all stardust (see Figure 1).

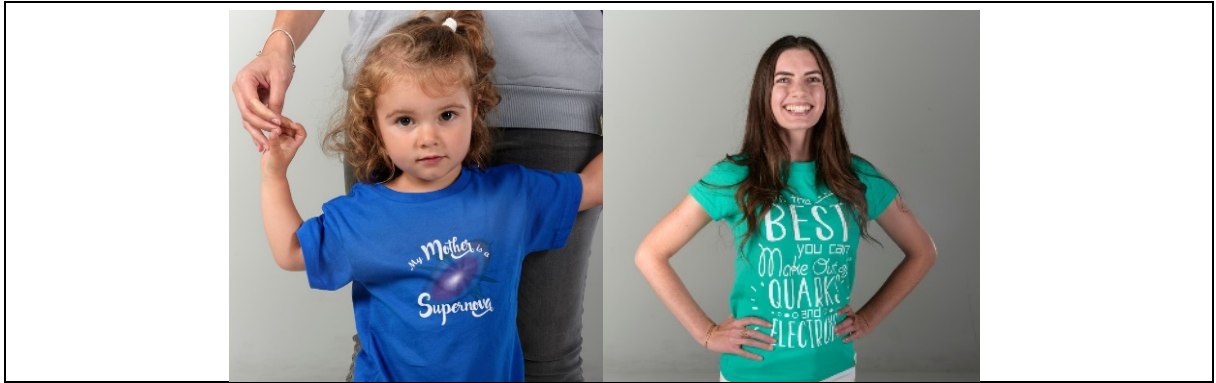

Figure 1: T-shirts from CERN's souvenir shop [9] stating "My mother is a Supernova" and "I am the best you can make out of quarks and electrons". All elements, except the three lightest, are products of reactions between atoms that take place inside stars. When a star dies in an explosion known as a supernova, heavier elements are thrown into space. Some have ended up on Earth over the last four billion years that our planet has existed. We are all stardust! For a more complete picture of the origins of atoms in the body see [10].

Particle physics can bring students access to new worlds and ways of thinking. When they access this microscopic world, students gain a better understanding of the structure of matter as well as that of new technologies. Particle physics gives them a new way to read the world around them, a fresh way to look at nature. If every child on this planet deeply understood the concept that our bodies are made of the same particles as everything else on the Earth and in the universe, that we are all part of Nature and everything is interconnected in the universe, greater respect towards Nature might be fostered.

Exposure to modern physics such as particle physics and its technological applications increases the interest of students in physics and their perception of today's role of physics and physicists in society. The impact of different types of extra-curricular activities in particle physics (e.g. exhibitions, the physics masterclass programme [11] and teaching [12]) on students aged 16-19 years in UK and Germany has been evaluated [13], clearly showing that their general interest in physics has increased very strongly. Students would like to deal more often with modern physics at school and criticized the lack of practical experiments.

Particle physics permits the demonstration of the scientific method, to show how advanced science is, where it is going, and how it works, and points out the technological advances in those processes. It also shows students how dynamic and collaborative research is, how experimentation has become both crucial and harder to accomplish. They learn the process through which science develops theories and validates models, how experiments are designed and built in a worldwide collaborative manner in order to validate theories and models through the analysis of data from large experiments.

Teaching more contemporary physics in schools may lead students to understand why so many investments, financial and personal, are made to develop laboratory facilities such as the LHC – one of the largest enterprises in mankind's history. Besides, it is important to discuss the experiments which are conducted at the LHC, so that some of the wrong information distributed by the communication media can be demystified, such as rumours of the possibility of the Earth being destroyed by a black hole created in the LHC, headlines about doomsday machine, etc.

### 1.3. Particle Physics at schools today

The importance of the efforts to broaden the spectrum of modern physics available at schools has been recently officially recognised by the full particle physics community. The European Strategy for Particle Physics Update 2020 recommended the teaching of the Standard Model in school classrooms: "The particle physics community should work with educators and relevant authorities to explore the adoption of basic knowledge of elementary particles and their interactions in the regular school curriculum." [14]

In the last decades scientists and educators started to develop materials suited to bring high school students in contact with the fascinating world of the fundamental building blocks of matter and their interactions, for example a plethora of resources developed by German Netzwerk Teilchenwelt, and requested by thousands of teachers all over Germany [12, 15, 16, 17] and other educational resources available for introducing the LHC in the classroom [18]. The only resources with broad reach to worldwide physics curricula are the International Baccalaureate (IB) Physics books [19, 20, 21] prepared for the purposes of IB Diploma Programme [22]. This is a two-year educational programme primarily aimed at 16-to-19 year olds, but IB offers four programmes for students aged 3 to 19. It was developed in the early to mid-1960s in Geneva, Switzerland, by a group of international educators. Today there are 7,002 IB programmes being offered worldwide, across 5,284 schools in 158 countries and nearly 30 000 students are assessed each year by the IB Organisation. If we add to this the 10 000 teachers trained at CERN since 2006 impacting more than a million of students [23], 4000 school students performing hands-on experiments at CERN's S'COOL LAB each year [24], 70 000 school children visiting CERN every year, 13 000 pupils in 60 countries yearly becoming scientists for a day through Physics Masterclasses [11] and many more participating in different activities (e.g. on cosmic rays [25, 26]) and even many enthusiastic teachers who make an effort to teach and promote extra-curriculum activities in particle physics, we would still end up with a tiny portion (smaller by two orders of magnitude) compared to world population of high-school age. This in big part explains the huge gap between science and society today, and the lack of awareness, understanding and trust of the general public and decision makers regarding science and its importance to society, leading to financial austerity for large scientific projects, low interest in the younger generations to become physicists, and a more general mistrust in science.

Albert Einstein was very clear on this subject: "It is crucially important that the general public has the opportunity to inform itself knowledgeably and intelligibly on the endeavors and results of scientific research. Restricting scientific findings to a small group of people weakens the philosophical spirit of a nation and leads to its intellectual impoverishment." (Princeton 1948).

## 1.4. Importance of outreach

It is for the reasons outlined above that current, well-focused concerted and global outreach and communication efforts to engage the public are already today of vital strategic importance.

Until we update our school curricula, science outreach and extra-curriculum activities are the only possibility to bridge the gap between science and society and bring a betterunderstanding of our fascinating world to schools and the public. IPPOG [5] has been making a concerted and systematic effort to present and popularise particle physics across all audiences and age groups for almost 25 years and today the scientific community has in IPPOG a strategic pillar for bringing contemporary physics and society closer.

# 2. IPPOG, a strategic pillar for particle physics outreach worldwide

In this chapter, IPPOG's mission and its different roles are introduced, namely of a network, an international collaboration, an exchange forum, a key partner for promoting scientific mission globally, and a strategic pillar to foster support for fundamental research.

## 2.1. IPPOG's mission

The International Particle Physics Outreach Group's principal aim is to maximise the impact of education and outreach efforts related to particle physics. It contributes to global efforts in strengthening the cultural awareness in the understanding and support of particle physics and related sciences, in raising scientific literacy in society, educating the public on the values of basic research and in developing and training the next generation of researchers, scientists and engineers.

In particular, IPPOG's purpose is to raise standards of global outreach and informal science education efforts of particle physics, to communicate its results and findings to the public, to bring new discoveries in all areas of particle physics research to young people and to convey to the public that the beauty of nature is indeed becoming understandable from the interactions of its most fundamental constituents - the elementary particles.

## 2.2. IPPOG's roles

IPPOG is a **network** of scientists, researchers, science educators, explainers and communication specialists active across the globe in outreach for particle physics. They come from prominent national or international professional physics centres, societies and laboratories engaged in particle physics research, and from major particle physics experiments. The diversity of their cultural and educational backgrounds brings a large and important variety of skills to the table, which permit for the effective development of novel outreach activities with maximal impact. IPPOG members represent links to several national-level science networks. This constitutes IPPOG's global network of laboratories, institutions, organizations and individuals all passionate about particle physics. The expertise of IPPOG's members spans all aspects of collider and non-collider research, including astroparticle physics and accelerator and detector technology.

As CERN's then Director General Prof. Chris Llewellyn Smith explained in September 1997 at the first-ever EPPOG (European Particle Physics Outreach Group, IPPOG's predecessor) meeting [27], "the particle physics community has a moral obligation to inform the public on its activities. To do this well, experiences must be shared among countries in view of the need to optimize the use of resources." He also highlighted the need of being selective on future actions and collective decision-making regarding outreach efforts, giving the example of informing school children as a good long-range investment. Out of this obligation to communicate effectively with the public in a targeted way and a need to pool resources, EPPOG was born. It was formed in 1997 under the joint auspices of the European Committee for Future Accelerators (ECFA) and the High Energy Particle Physics Board of the European Physical Society. EPPOG widened its regional scope to become an international player in 2005 with the development of the International Particle Physics Masterclass programme [11], and became officially renamed as the International Particle Physics Outreach Group (IPPOG) in 2011.

With the growing global scale of IPPOG activities, and taking into consideration recommendations of the 2013 European Particle Physics Strategy Update, IPPOG became an **international collaboration** based on a memorandum of understanding in 2016. The signature for each country is provided by a single entity (e.g. a ministry, lead scientific institution, or in some cases a university that has a national leading role) that oversees or coordinates the efforts of particle physics outreach in that country. The person signing for an international experiment is typically the collaboration spokesperson or a member of management responsible for the outreach programme. CERN's signature comes from the head of International Relations. These bodies each select representatives who are identified as main actors in the field of particle physics outreach.

Today the IPPOG Collaboration comprises 33 members: 26 countries, 6 experiments and CERN as an international laboratory. Thanks to its ever-growing membership and global coverage, IPPOG fosters the recognition and raises the value of science outreach around the world. It increases international exposure, enables centralised and coordinated efforts through partnerships.

IPPOG is also a **forum** for the exchange of information and best practices with colleagues from around the world, a brainstorming platform, a source of ideas and inspiration, a ground for training and skill-development (for public talks etc.), a platform providing access to programmes for schools, and a place to find partners for common outreach and education projects and publications.

In addition to its bi-annual meetings, IPPOG forum members are active in different working groups focusing on specific issues and developing best practices to be shared within IPPOG and other outreach and education forums and experts around the globe. These are for example 'Bringing Masterclasses to New Countries', 'Explaining Particle Physics Hot Topics to a Lay Audience', 'Outreach of Applications for Society' and 'Exhibits and Public Events'.

For particle physics and scientific community IPPOG represents a **key partner** for promoting and enabling their scientific mission and activities globally, a platform for engagement on a global level, building partnerships within the community and across communities, and for supporting the broader scientific objectives of particle physics and its role in the society on global level. The worldwide particle physics community has in IPPOG a strong partner at hand when reaching out to the wider society in diverse ways that are adapted for every target audience.

IPPOG is a **strategic pillar** for future of particle physics and scientific community and is helping to foster long-term, sustainable **support for fundamental scientific research** around the world. IPPOG develops programmes and strategies to address the current and future challenges of the particle physics and scientific community, such as the declining interest in STEM-related studies, lack of support of fundamental research and mistrust in science. IPPOG's scientific education and outreach activities aim to improve public understanding and appreciation of the benefits of fundamental research, to spark interest and enthusiasm among young people, and to strengthen the integration of science in society. IPPOG not only motivates, inspires, and educates our youth in the field of particle physics, but it develops personal awareness of the value of science and of the process of evidence-based decision making, regardless of age or discipline. IPPOG helps to establish broad public support, as well as the commitment of key stakeholders and policy makers throughout Europe and the world for the future large-scale projects of particle physics community. More details on IPPOG's strategic mission in tackling the future challenges in particle physics education and outreach both in the context of the European Particle Physics Strategic Update (EPPSU) and in general for our society are described in [28].

In a confusing world of endangered reasoning, IPPOG works to strengthen the **trust in science** and its method of **evidence-based decision making** to offer future generations a meaningful basis that generates supportive structure in their life. Without compromising established methods, IPPOG is exploring new paths to engage citizens – especially the young. Reaching out to high-school students and their teachers to convey the methods and tools used in fundamental science is a strong investment in the future. While only a fraction of young students will become scientists, and fewer still will become particle physicists, all will become ambassadors for the scientific method and evidence-based decision-making. Younger audiences will be more educated and appreciative of the importance of research, and thus more suited to make informed decisions about science and scientific questions for their nations and their peoples in the future [29].

## 2.3. IPPOG Activities

Current IPPOG activities include the ever-growing and well-established International Masterclasses on Particle Physics programme [11], the outreach Resource Database (see chapter 3)[37], the Global Cosmic Rays experiments at schools platform currently in development [30], support for exhibitions and activities at public events and festivals (e.g. Colours of Ostrava in Czech Republic [31], Universal Science [32]), and different topical programmes and competitions targeting young and diverse audiences (like Particles4U [33], Girls, do Physics! [34], Creating Ambassadors for Science in Society [4], Cascade competitions in UK [35] and Slovakia [36] etc…). These diverse activities allow IPPOG to bridge the gap between science education at school and modern scientific research by inspiring, motivating and educating an especially young audience, offering hands-on experience and connecting physics to real life while using cutting edge technologies.

### 2.3.1. Physics Masterclasses

The Particle Physics Masterclasses programme [11] has been IPPOG's flagship activity for the last 15 years. Research institutes and universities around the world invite students and their teachers for a day-long programme to experience life at the forefront of basic research. The programme gives students the opportunity to become a particle physicist for a day by analysing real data collected by the experiments at the CERN LHC and other particle physics facilities (e.g. Japanese Belle II, American MINERvA neutrino or particle cancer therapy masterclasses). The International Masterclasses (IMC) run every Spring and attract each year more than 13 000 high school students in 60 countries. Participants can explore the fundamental forces and building blocks of nature and are informed about the new age of exciting discoveries in particle physics, e.g. the discovery of the Higgs boson. Moreover, they can actively get a flavour of cutting-edge research and improve their understanding of science and the scientific research process while mirroring exactly the activities of working particle physicist. The programme offers authentic experience and is a valuable addition to physics education at school, thus stimulating the students' interest in science [7]. At the end, like in a real international research collaboration, the participants join a videoconference hosted at CERN or Fermilab to combine and discuss results and ask questions. The programme continues to broaden; special versions of IMC were developed recently, such as the International Day of Women and Girls in Science Masterclasses (IMC engaging young women in science) and Worldwide Data Day.

Moreover, the physics masterclasses website [11] offers a collection of educative materials on particle physics, which includes interactive elements, for example students' exposure to real particle physics experimental data allowing students to make their own measurements, and understand particle physics "hands-on".

## 3. IPPOG Outreach Resource Database

The richness of IPPOG's expertise given the diversity of its members in terms of cultural and educational backgrounds, provides a perfect platform for sharing, developing and improving the explanatory and teaching materials, strategies, methods, activities and tools to reach broader audiences, based on the best practices in outreach and education of particle physics and related sciences. In its commitment to foster science dissemination, an online platform has been built to facilitate the exchange of particle physics education and outreach resources across the globe. IPPOG's Resource Database (RDB) [37] is a collection of high quality engaging materials (e.g. videos, posters, talks, hands-on activities, tools, brochures and more) recommended by IPPOG representatives and contributors to help sharing the wonders and excitement of particle physics with teachers, students and the general public. The information available is readily understandable and regularly updated to reflect the latest discoveries in particle physics. Everything is freely available with the spirit of open access. The items are submitted to the database by IPPOG representatives, members of IPPOG Forum and group of trusted contributors. Anybody can request to become a new contributor and will be added to the group upon the approval from the IPPOG chairs.

### 3.1. IPPOG Resource Database history

The idea of the RDB was born in 2009 at an EPPOG meeting at CERN, when EPPOG after more than 15 years of its existence was on the verge of a transformation from a "discussion forum" to a possible world leader in outreach and informal scientific education activities for particle physics and related fields. Initially called the EPPOG Best Practice Database, it was meant to be used by science institutions and laboratories for outreach and informal science education purposes. The eventual aim was that the database would be self-sustaining, with the EPPOG members (and others) adding new items when appropriate and users "voting" to ensure that the highest quality and most useful items are the ones that become the "best practice" [38]. Consequently, the first version of the IPPOG's Resource Database was released in 2011.

### 3.2. New IPPOG website

After almost 10 years, IPPOG is now embarking on an ambitious project to improve the user experience across the IPPOG digital portfolio (website and social media channels) and to strengthen the IPPOG brand online by creating a new website including a new RDB. The goal of the new design is to greatly broaden the audience type and use of the web pages and available resources. The visual impact needs to be enhanced in order to pursue IPPOG's mission and communicate its messages to its existing and new audiences, as well as new potential members, partners and sponsors. IPPOG wants the new website to become more open to students, teachers and the general public, and for the RDB to become the primary source of particle physics outreach material in the world [39]. Comparison of current and new designs of IPPOG website are shown on Figures 2 and 3. Entry to RDB is possible from the main menu, but it is made prominent also at the front page (Figure 3).

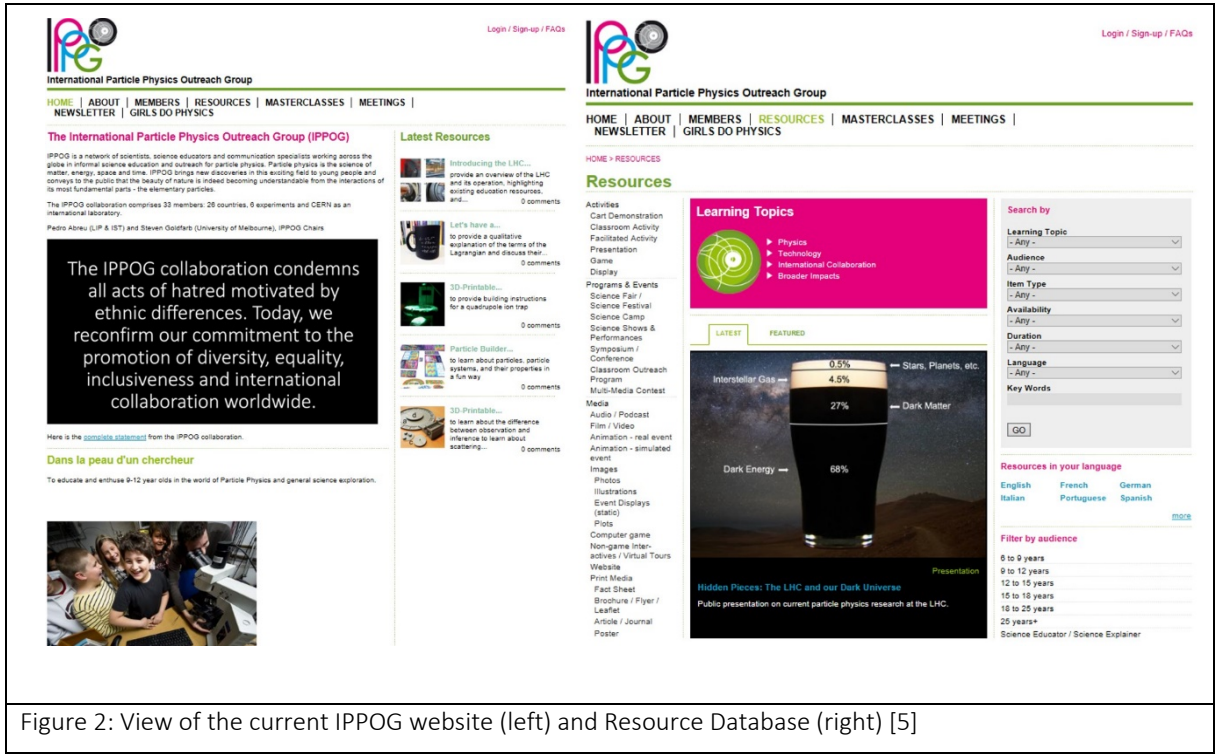

Figure 2: View of the current IPPOG website (left) and Resource Database (right) [5]

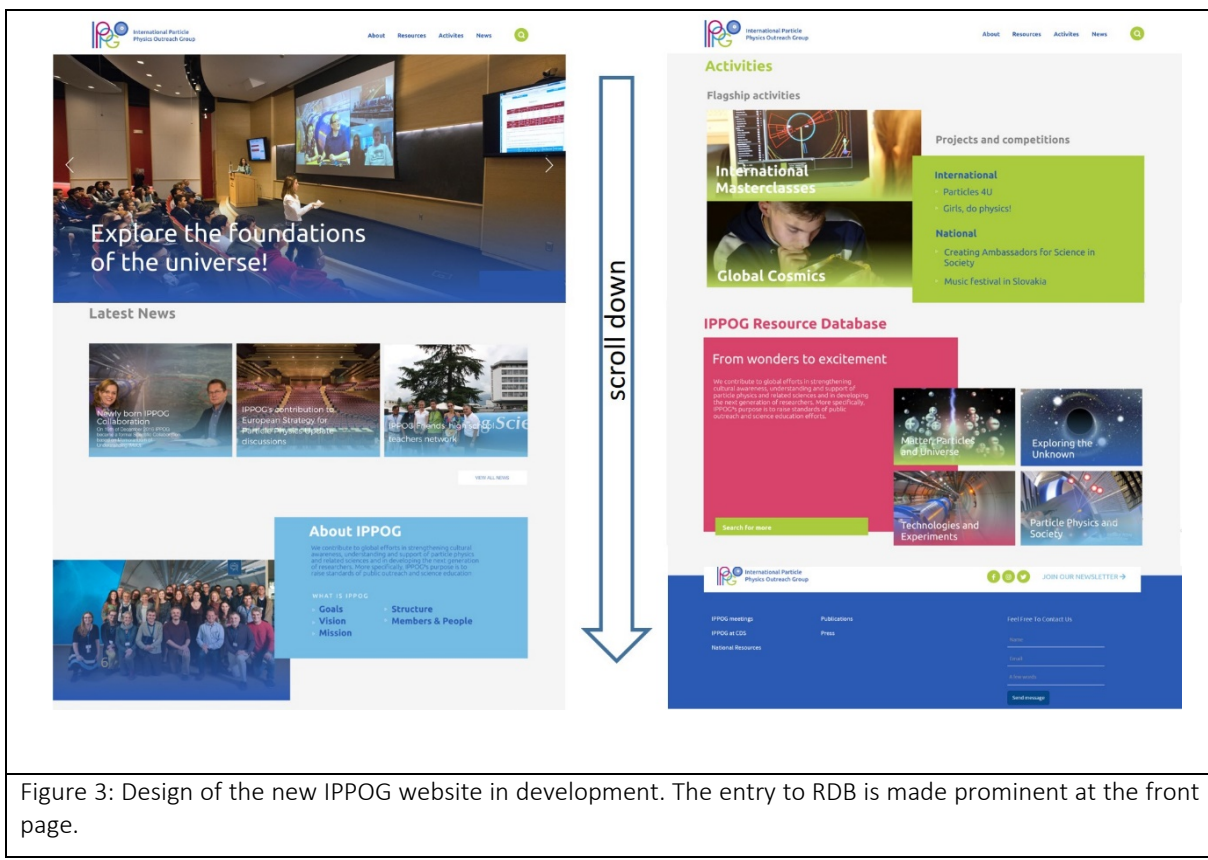

Figure 3: Design of the new IPPOG website in development. The entry to RDB is made prominent at the front page.

Considerable efforts are invested in the review and redesign of the RDB, which will also benefit from the visual upgrade (see Figure 4).

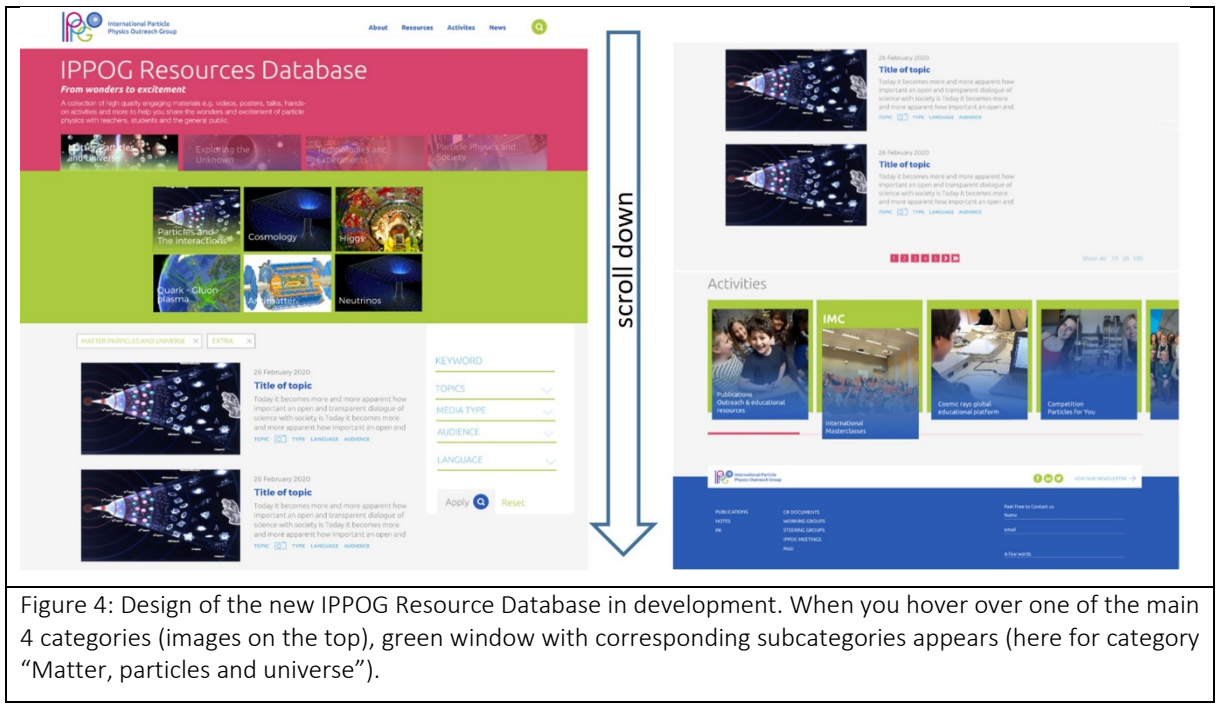

Figure 4: Design of the new IPPOG Resource Database in development. When you hover over one of the main 4 categories (images on the top), green window with corresponding subcategories appears (here for category "Matter, particles and universe").

### 3.3. The new IPPOG Resource Database

Currently there are almost 400 items in RDB which are filtered by topics, item types, target audiences and language. Users can choose among 44 physics topics (from particles and their interactions through dark matter, extra-dimensions to detectors and accelerators), 41 item types (e.g. video, poster, lesson plan, etc.) (see Figure 5 and [37]).

For several years discussions within IPPOG were considering how to improve the functionality, keep the content up-to-date and provide the full range coverage of relevant physics topics. As a result of the IPPOG Web working group efforts over many years and numerous consultations with science communication experts and representatives of the IPPOG target audiences the proposal of a new RDB has been developed [39]. A large number of categories, item types and audience types have been collapsed to fewer more targeted ones (see Figure 5) and a more user-friendly way of navigating and search was proposed [39]. These have been approved by the IPPOG community and teachers. As shown on Figure 4, when the RDB page is entered, one would see images on the top corresponding to the main four physics topics, while the search filter is located below on the right hand side. When one hovers over the one of images, the green window with images of the corresponding subcategories appears. The main idea is that one can search by a filter, but also use the quick search by topics by clicking directly on the images on the top, while at every step there is a possibility to refine the search using the filter on the side, which always remains in view.

**IPPOG RESOURCE DATABASE NEW/CURATED CATEGORIES**

**SEARCH FILTERS**
- TOPIC / SUBTOPIC
- ITEM TYPE
- AUDIENCE
- LANGUAGE
- SCHOOL TOPIC
- ONLINE USAGE
- KEYWORD
- IPPOG'S BEST

**TOPICS**

1) **MATTER, PARTICLES AND UNIVERS (KNOWN PHYSICS)**
PARTICLES
INTERACTIONS
COSMOLOGY
HIGGS
ANTIMATTER
QUARK-GLUON PLASMA
NEUTRINOS

2) **EXPLORING THE UNKNOWN (BEYOND KNOWN PHYSICS)**
SUPERSYMMETRY
DARK MATTER
DARK ENERGY
EXTRA DIMENSIONS

3) **TECHNOLOGIES and EXPERIMENTS**
ACCELERATORS
DETECTORS

4) **PARTICLE PHYSICS AND SOCIETY**
WHY FUNDAMENTAL RESEARCH
INTERNATIONAL COLLABORATION
APPLICATIONS & SPIN-OFFS
PEOPLE BEHIND THE SCIENCE

**ITEM TYPES**
PHOTOS / POSTERS / CHARTS
VIDEOS
ANIMATIONS / SIMULATIONS
PRESENTATIONS / TALKS
GAMES
CLASSROOM MATERIALS / TUTORIALS / LESSON PLANS / TEXT BOOKS
BOOKS
PROJECTS / COMPETITIONS
EXHIBITIONS / EXHIBITION ITEMS
SOUVENIRS

**AUDIENCES**
PRIMARY SCHOOL LEVEL
LOWER SECONDARY SCHOOL LEVEL
UPPER SECONDARY SCHOOL LEVEL
BROAD PUBLIC
EDUCATORS

Figure 5: New categories of IPPOG Resource Database.

Taking into consideration that teachers and educational specialists are the primary audience of RDB, since 2017 IPPOG has worked closely with two groups of physics teachers attending the international high school teachers programme at CERN [23]. They reviewed and evaluated a large number of materials, proposed new tags, helped to determine the structure, interface and workflow. A nice "spin-off" from working with teachers was the creation of the group "IPPOG Friends", a group of physics teachers interested to learn about IPPOG related activities, take part and disseminate them among their colleagues and students.

Nine years after the creation of the RDB, IPPOG has created an official RDB curation group, including teachers, scientists and science educators from 15 countries, 3 experiments and CERN. They are reviewing the RDB content and reorganising it according to new categorisation. Many new items are being collected at the same time in order to provide the most comprehensive and up-to-date collection of outreach and educational materials in particle physics and related sciences worldwide.

Even though most of the resources are in English, thanks to the large international representation of IPPOG, the materials spans 24 languages. In the future we aim to work on translation of the most recommended materials to various languages. The curation group will also identify IPPOG's best resources that will always appear at the top of the search results. Based on the request from teachers, a new tag 'school topic' is added, which helps teachers to identify where in their classical physics curriculum they can use the particular resource. The creation of the list of school classroom topics linking to respective particle physics topics is a very useful tool being now developed by IPPOG. Another new tag is the possibility to use the resource also for online educating.

## 3.4. Target audiences (who can find what?)

The first primary group of IPPOG RDB users (particle physics and scientific community active in outreach and education) are expected to seek the materials in support of their own particle physics outreach projects and inspiration (e.g. for public talks). The second primary group, teachers and educators, are expected to find the extra-curricular activities, lesson plans or projects for the classroom, and even the materials to inspire young students and motivate them to study STEM subjects and become scientifically aware citizens. Curious students will find materials or activities in particle physics and related sciences complementary to what they are being taught at school.

Good examples of useful resources for the high school physics teacher are instructions how to build a simple cloud chamber [40], a 3-D printed scattering experiment [41], a Particle Builder board game [42] or a multilingual poster describing the elementary constituents of matter [43]. A four-piece series of particle physics teaching materials [15], and in particular the main volume "Charges, Interactions and Particles" [16, 17] which covers the essential knowledge about the Standard Model by using an innovative approach tailored around the concept of charge and pointing out connections to typical contents of school physics curricula, is very popular with thousands of German teachers. Even primary school teachers can find an inspiration, e.g. with CERN training programme Playing with Protons [44].

If you are a scientist invited to give a public talk on the Higgs boson and other intriguing questions in particle physics, you might be inspired by different talks [45], presentations and other materials, e.g. the Higgs boson comic book and video animation [46]. When you are confronted by media or people with negative attitude towards science, you might want to look at the presentation discussing the impact of fundamental research on society showcased by particle physics [47].

### 3.5. Bringing particle physics closer to society

Particle physics has not only revolutionized the way we look at the universe, but along the way, it has made significant impacts on other fields of science. As a key driver of innovation through its countless applications, it has improved the daily life for people around the world, showing particle physics is a science in service of mankind. A new generation of scientists and computing professionals has been trained, and science for peace initiatives led especially by CERN are representing the DNA of the open collaborative spirit within particle physics.

It is important to involve also these lesser known aspects of fundamental research in teaching and outreach activities. "*Many people are not aware that behind all innovations, there is science. I wasn't aware of most of the inventions coming from CERN, which are currently essential for our everyday lives.*", claim high school students attending the Creating Ambassadors for Science in Society event [4]. Understanding what is the direct impact of particle physics and fundamental research on everyday life is important in shaping the perception and relation to physics and science as such and even a motivating factor for undertaking physics studies: "*We understood that physics plays a relevant role in today's technologies and it will do so even more in the future. Its applications, such as hadron therapy among others, fascinated and encouraged us to study and continue our college education up to completing a Ph.D.*" as shown in the testimonial from winners of 'Girls, do physics!' IPPOG's campaign [34].

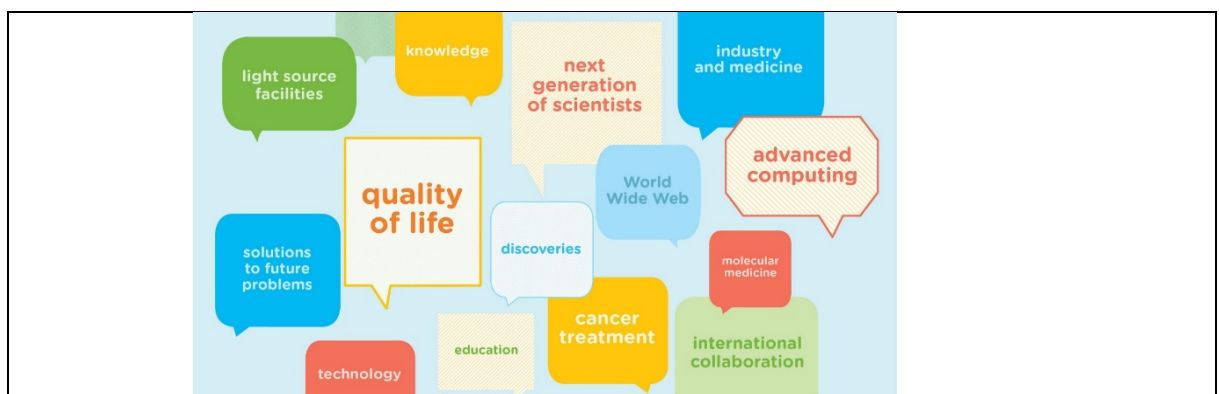
Figure 6: Why particle physics matters [48], Illustration by Sandbox Studio, Chicago

In this direction, IPPOG offers today for example a Masterclass on hadron therapy [49] and a small number of other resources [47], but many more are being prepared.

An important new part of the RDB will be the category "Particle physics and society", which will be filled with new useful tools for our audiences which are being prepared thanks to the efforts of the IPPOG working groups.

The working group on "Outreach of Applications for Society" identified the need for engaging stories with a human touch since they can be the most impactful and effective way to communicate to the public successful applications for the benefit of society from fundamental research and physics in particular. Its aim is to offer short stories with clear messages connecting science to everyday lives of citizens and make them available through the IPPOG web page and resource database. The goal is to also show through them the importance of fundamental research and related sciences for global sustainable development and to inspire the young generation. These stories should provide a powerful tool when approaching different non-scientifically educated citizens realizing that in addition to the quest for knowledge and satisfying curiosity, there is a growing pressure from taxpayers for the justification of fundamental research funding requesting tangible examples of return to society.

After collecting materials, from the CERN knowledge transfer collection to individual contributions, the working group works on story-building around available facts, while trying to find a human touch and anecdotes in each one of them. Almost everybody knows the story behind the World Wide Web. We plan to offer more stories, behind many applications, working currently on subjects like PET and IRM scanners, cancer therapy, touch screen, GPS, UNOSAT and many more little known even within the scientific community.

Another IPPOG working group is regularly discussing how to explain complex particle physics hot topics to lay audience. The highlights of the discussions in recent years include how to best explain the Higgs mechanism, spin, particle-wave duality or gravitational waves. However, also questions such as "how to communicate the importance of precision", "what if no new discovery / Higgs – what now?", "what is the meaning of theory in science compared to its meaning for the broad public at large" or "how to sell a new European Strategy for Particle Physics and how to justify the costs of always-bigger machines (HL-LHC, ILC, FCC)" have been tackled. In 2019 work started on collecting the best IPPOG recommended explanations, analogies, metaphors and examples on how to explain complex particle physics issues to the public. All the categories in IPPOG's resource database are planned to be addressed. Arguments that can be used when talking to media or decision makers, when they question the funding of research projects and its relevance, will be included in the "Particle Physics and Society" part of the RDB. The scientific outreach community and everybody who wants to explain these issues to the public, students and politicians or to convince media and decision makers why particle physics and fundamental research matters, will find this to be a powerful tool.

# Conclusions

Threatened financial support for large experimental endeavors, falling interest of young people to engage in studies of STEM, especially physics, and mistrust in science are the main challenges the scientific community is faced with currently. These are based on the misperception of science, especially physics and basic research in society. This attitude towards physics by non-scientific audiences is largely caused by the scarce exposure of society to modern physics, which is in most cases not included in school curricula. Introducing particle physics and related sciences to students and the public, while showing the current state of art of contemporary physics and bringing the understanding of the world we live in and its technologies, is of vital importance. Before the update of physics school curricula is achieved, the science outreach and extra-curriculum activities are the only possibility to bridge the gap between contemporary scientific research and science education at schools and thus the awareness of modern science by society. The International Particle Physics Outreach Group (IPPOG) has been making concerted and systematic efforts to present and popularise particle physics across all audiences and age groups for more than two decades by developing suitable methods, tools and activities. Today the scientific community has in IPPOG a strategic pillar to aid the fostering of long-term, sustainable support for fundamental research around the world. One of the main tools IPPOG has been offering the scientific community, teachers and educators for almost 10 years is the Resource Database (RDB), an online platform containing a collection of high quality engaging education and outreach materials in particle physics and related sciences. Today IPPOG is investing considerable efforts in redesigning the RDB with the aim for it to become the primary source of particle physics outreach material in the world. At the same time, new content for the RDB is being prepared, with the aim to provide a powerful tool for the scientific community to shape the attitude and perception of physics and fundamental research by decision makers, funding bodies and the public, and to motivate young people to undertake physics studies.

## Acknowledgement

Author wishes to express gratitude to Dezso Horvath for encouragement, comments and help in producing the final version of this article.